\documentclass[conference]{IEEEtran}
\usepackage[T1]{fontenc}
\usepackage{amsmath,amsthm,amssymb,booktabs,graphicx,cite}
\usepackage{tikz}
\usetikzlibrary{positioning,arrows.meta}
\newcommand{\tkdrop}{6mm}
\tikzset{factory/.style={font=\footnotesize, >=Stealth, node distance=10mm and 5mm},
  box/.style={draw, rounded corners=2pt, align=center, minimum height=8mm, inner sep=3pt},
  mem/.style={box, fill=black!6}, hum/.style={box, double},
  lbl/.style={font=\scriptsize, align=center},
  comp/.style={text width=34mm, minimum height=14mm},
  compgrid/.style={node distance=9mm and 16mm}}
\IfFileExists{tables/numbers.tex}{\newcommand{\ratioFirst}{0.40}
\newcommand{\ratioLast}{1.50}
\newcommand{\nameFirst}{stock\_game}
\newcommand{\nameLast}{learning\_agent}

\newcommand{\nApplied}{36}
\newcommand{\nRejected}{0}
\newcommand{\pctApplied}{41}
\newcommand{\nFInRatio}{88}
\newcommand{\nIncreases}{3}
\newcommand{\nSteps}{4}
\newcommand{\firstRuleDate}{2026-11-08}
\newcommand{\lastRuleDate}{2026-11-27}
\newcommand{\nDaysObserved}{20}
\newcommand{\nFlows}{18}
\newcommand{\maxFlow}{17}
\newcommand{\maxFlowPairs}{quant\_home to template and quant\_home to learning\_agent}
\newcommand{\nOriginTemplate}{14}
\newcommand{\nRepos}{6}

\newcommand{\nNativeRules}{104}
\newcommand{\nPortedRules}{123}
\newcommand{\nProposals}{96}
\newcommand{\cutoffDate}{2026-11-27}
\newcommand{\nUniqueRules}{104}
\newcommand{\nPortedCopies}{123}
\newcommand{\nRecur}{5}
\newcommand{\pctRecur}{4.8}
\newcommand{\nRecurRecipient}{0}
\newcommand{\nCheckRules}{55}
\newcommand{\nProseRules}{49}
\newcommand{\nRecurCheck}{5}
\newcommand{\nRecurProse}{0}
\newcommand{\pctRecurCheck}{9.1}
\newcommand{\pctRecurProse}{0.0}

\newcommand{\nCopiedNotes}{3}
\newcommand{\nGateDefects}{10}
\newcommand{\nGateBySecondAgent}{8}
\newcommand{\nGateByProduction}{1}
\newcommand{\nGateByUAT}{1}
\newcommand{\nGateByHarness}{0}

\newcommand{\topClass}{Tests and gates}
\newcommand{\nTopClass}{38}
\newcommand{\nOtherClass}{12}
\newcommand{\nRecurClasses}{3}
}{}
\newtheorem{definition}{Definition}
\newtheorem{proposition}{Proposition}
\newcommand{\cA}{C1}
\newcommand{\cB}{C2}
\newcommand{\cC}{C3}
\newcommand{\cD}{C4}
\newcommand{\cE}{C5}
\newcommand{\rqA}{RQ1}
\newcommand{\rqB}{RQ2}
\newcommand{\rqC}{RQ3}
\newcommand{\rqD}{RQ4}
\title{From Software Factory to Solution Factory: Bounded, Human-Gated Recursive Self-Improvement of an LLM
Agent Pipeline Across Projects}
\author{\IEEEauthorblockN{Qiyu Zhang}
\IEEEauthorblockA{School of Information Technology, Nanyang Polytechnic, Singapore\\
Email: joey\_zhang@nyp.edu.sg}}
\begin{document}
\maketitle
\begin{abstract}
Recursive self-improvement of AI agents is usually studied as open-ended self-rewriting scored by a benchmark.
We study a bounded form in production: an LLM software factory that turns its failures into written rules,
carries them across projects, and changes its own scripts only through proposals a human approves. Across
\nRepos{} repositories it accumulated \nNativeRules{} rules, ported \nPortedRules{} copies between
repositories and received \nProposals{} improvement proposals. We tested four claims against tests fixed in
advance. No ported rule recorded a recurrence in its new repository, while \nRecur{} of \nUniqueRules{} rules
recorded one where they were written: weak, self-reported support for transfer. Rules that named a mechanical
check recurred at least as often as prose rules; a short Poisson model shows how this comparison can
be confounded by how often the guarded failure occurs. The deterministic gates guarding the loop had defects their
own harness missed: of \nGateDefects{} gate defects, the harness found \nGateByHarness{} and an independent
second agent found \nGateBySecondAgent{}. Proposals per delivered task, our proxy for human load, did not fall:
they went from \ratioFirst{} to \ratioLast{} across \nSteps{} project-to-project steps. The same loop built a non-software pipeline, the one
that wrote this paper, which we present as a step from a software factory to a solution factory.
\end{abstract}
\section{Introduction}
\label{sec:intro}

Recursive self-improvement (RSI), in which a system modifies the machinery that produces its own future
behaviour, has moved from a theoretical concern to an engineering practice. Coding agents now rewrite their own
scaffolds and retain the variants that score higher on a benchmark
\cite{zhang2025darwin,robeyns2025a,wang2025huxleygodel,liu2026mendel}, and research agents optimise the code
that runs their own experiments \cite{srikanth2026recursive}. In this line of work the fitness signal is a
benchmark score, and the improvement loop runs without a human in it.

Teams that deploy coding agents in daily work operate a different loop. When an agent fails, the failure is
written down together with its cause and a rule for avoiding it, and subsequent runs read that record. We refer
to this as \emph{bounded} RSI: the system does improve itself, but every change to its own machinery is a
written proposal that a human accepts or rejects, and every lesson is a human-readable rule that, where
possible, names a mechanical check. Bounded RSI is less ambitious than open-ended self-rewriting. It is,
however, the form that is deployed today. Its first measurements have appeared recently: a failure ledger kept
over six weeks in one repository \cite{aliabadi2026lessons}, and a governed loop that admits learned rules on
two repositories \cite{heander2026admitted}. Both study rules within the repositories where they arose. What
happens to a rule once it is carried to another project, and whether the gates that guard such a loop are
themselves sound, has not been measured.

This paper reports such a measurement. The object of study is a \emph{software factory}: a set of shell
scripts, prompts and memory files with which large language model (LLM) agents turn a specification into merged
and tested code. Across \nRepos{} repositories of one lineage, observed over \nDaysObserved{} days, the factory
accumulated \nNativeRules{} pitfall rules, propagated \nPortedRules{} copies of them between repositories and
received \nProposals{} proposals to change itself. Because these records are kept in version control, the
creation, propagation and recurrence of each rule can be reconstructed. The pipeline that produced this paper
was itself built by the factory. We take this as a first instance of what we call a \emph{solution factory}:
the same improvement loop applied to producing gated pipelines for work other than software.

We address four research questions, each with a test fixed before the analysis.
\begin{itemize}
\item \textbf{\rqA{} (transfer, \cA{}).} Do rules learned in one project carry over to the projects that receive
them?
\item \textbf{\rqB{} (rule form, \cB{}).} Do rules that name a mechanical check recur less often than rules
stated only in prose?
\item \textbf{\rqC{} (verification, \cC{}).} Do the deterministic gates that guard the loop themselves fail, and
who detects such failures?
\item \textbf{\rqD{} (human cost, \cD{}).} Does the human effort per delivered task decrease as the factory
accumulates rules?
\end{itemize}
A fifth claim (\cE{}) is a position on the extension from software to other products.

The findings are mixed. No ported rule recorded a recurrence in its receiving repository
(\nRecurRecipient{} of \nPortedCopies{} copies), which is consistent with transfer but rests on self-reported
recurrence and is therefore weak evidence. Contrary to expectation, all \nRecur{} recorded recurrences concern
check-backed rules; a short Poisson model shows how this comparison can be confounded by failure frequency. Of
\nGateDefects{} defects found in the factory's own gates, its test harness detected \nGateByHarness{}, whereas
an independent second agent detected \nGateBySecondAgent{}. Proposals per delivered task did not fall; they
rose from \ratioFirst{} to \ratioLast{}.

The contributions are:
\begin{enumerate}
\item an operational definition of bounded, human-gated RSI and a description of a factory that implements it
(Sections~\ref{sec:factory} and~\ref{sec:model});
\item to our knowledge, the first measurement of rules propagated between repositories: their creation,
porting and recurrence across \nRepos{} repositories, pinned to fixed commits (Sections~\ref{sec:design}
and~\ref{sec:results});
\item a count of defects in the loop's own deterministic gates by who found them, as evidence that the
verifiers of such a loop require independent verification;
\item a report of two negative results, with a model that explains why one of them cannot be read causally.
\end{enumerate}
The study is small: a single operator, \nDaysObserved{} days and \nRepos{} related repositories. Section~\ref{sec:discussion}
discusses these limits.

\section{Background and Related Work}
\label{sec:related}

Table~\ref{tab:compare} positions this work against representative approaches along five dimensions. Each
entry reflects only what the cited abstract states.

\begin{table*}[t]
\centering
\caption{Representative approaches to agent self-improvement, as described in each abstract. ``Human gate'': a
human approves each change to the improving artefact. ``Across'': lessons are carried between unrelated
codebases or tasks. ``Natural language'': the improved artefact is prose rather than code. n.s.: not stated in
the abstract.}
\label{tab:compare}
\resizebox{\textwidth}{!}{%
\begin{tabular}{llllll}
\toprule
Approach & What is improved & Improvement signal & Human gate & Across & Natural language \\
\midrule
Darwin G\"odel Machine \cite{zhang2025darwin} & agent's own code & coding benchmark & n.s. & no & no \\
Self-improving coding agent \cite{robeyns2025a} & agent's own code & benchmark tasks & no (autonomous) & no & no \\
AIDE$^2$ \cite{srikanth2026recursive} & research agent's code & hidden evaluations & n.s. & held-out benchmarks & no \\
SpecForge \cite{cheng2026specforge} & behavioural specifications & execution-verified tests & n.s. & no & yes \\
Reflexion \cite{shinn2023reflexion} & reflections in memory & task feedback & n.s. & no & yes \\
ExpeL \cite{zhao2023expel} & extracted experiences & task outcomes & n.s. & across tasks & yes \\
Voyager \cite{wang2023voyager} & skill library & environment feedback & no & across tasks & no (code) \\
Accumulated rules \cite{aggarwal2026selfimproving} & rule file, checklist & accepted review comments & yes & one platform & yes \\
Retrospective harness changes \cite{treude2026how} & harness & later issues & by researchers & same repositories & yes \\
Failure ledger \cite{aliabadi2026lessons} & rule set, checks & recorded failures & n.s. & one repository & yes \\
Notary-gated retrospective \cite{heander2026admitted} & learned rules & review comments & agent notary & two repositories, separately & yes \\
This work & rules and factory scripts & exit codes, recorded failures & yes & \nRepos{} repositories & yes \\
\bottomrule
\end{tabular}}
\end{table*}

\paragraph{Self-rewriting agents}
The Darwin G\"odel Machine maintains an archive of coding agents that modify their own code and selects among
them by benchmark performance \cite{zhang2025darwin}. Subsequent systems refine the selection
\cite{wang2025huxleygodel,liu2026mendel} or allow a single coding agent to edit its own scaffold
\cite{robeyns2025a}. AIDE$^2$ applies the loop to a research agent that proposes changes to its own code and
retains those that perform best on hidden evaluations \cite{srikanth2026recursive}. Surveys and benchmarks now
organise the area \cite{ren2026selfimprovements,ibrahimoglu2026autonomous,zhang2026flexrsi,chi2026ai4aibench}.
These systems are open-ended and benchmark-scored; ours is human-gated and scored by delivered work. The
distinction is one of design, and no performance comparison is attempted.

\paragraph{Improving context rather than code}
A model with frozen weights can improve by refining what it reads. SpecForge refines behavioural
specifications using execution-verified tests \cite{cheng2026specforge}, and several systems evolve the agent
harness \cite{wu2026modularrsi,shang2026meshharness,dong2026mixture,karten2026continual,xia2025livesweagent}.
These operate within one repository or benchmark, whereas the rules studied here travel between unrelated
projects.

\paragraph{Experiential memory}
Reflexion stores verbal reflections on failed attempts and conditions later attempts on them
\cite{shinn2023reflexion}; ExpeL extracts experiences and insights from past tasks for reuse
\cite{zhao2023expel}; and Voyager accumulates an ever-growing library of executable skills
\cite{wang2023voyager}. Experiential co-learning transfers shortcuts between software-developing agents
\cite{qian2023experiential}, cross-task experiential learning reuses experience in multi-agent collaboration
\cite{li2025crosstask}, and reflective learning has been proposed for agents in specialised environments
\cite{allard2026experiential}; Evo-Memory benchmarks memory that evolves at test time \cite{wei2025evomemory}.
In this line, memory is consumed by the agent itself. In the factory, the memory also constrains the
orchestrator and is curated by a human, which is closer to the lessons-learned practice of software
organisations \cite{bose2008lessons}.

\paragraph{Rules learned from mistakes}
The two closest studies are recent. A six-week case study of one repository edited daily by three agent tools
under one shared rule set keeps a ledger of 118 agent failures, their fixes and whether they came back, and
finds that stale documentation was the largest failure class \cite{aliabadi2026lessons}. A pre-registered study
of 880 pull requests on two Python repositories runs a retrospective loop in which an agent proposes edits to
learned rules and a notary agent admits or rejects them; 55 of 89 proposals were admitted, but held-out review
comments and hidden-test resolution did not improve \cite{heander2026admitted}. Both measure rules inside the
repository where they arose, and in the second the gate is an agent. Here rules are copied between
repositories, the gate is a human, and the gates themselves are audited. Earlier work codifies each accepted review comment as a persistent rule in a version-controlled
instruction file and reports no recurrence of the ruled-against error classes on one microservices platform
\cite{aggarwal2026selfimproving}. Validation is the concern of a study of harness self-evolution on SWE-bench
issues: only some implemented changes reduced the targeted behaviour \cite{treude2026how}. PRECEPT reports that coding agents reintroduce a vulnerability class they had fixed once a
new session begins, and evolves security principles to prevent it \cite{guo2026precept}; a dual memory has been
applied to testing and bug localisation \cite{mo2026selfevolving}. Our recurrence measure differs from these:
it consists of dated notes, across projects, written by the agents and the operator, and it is not comparable
to a zero recurrence rate on one platform. The added element is structural: rules that govern the factory
rather than one agent's coding habits, propagated across projects, with self-changes gated by a human.

\paragraph{Verifying the verifier}
Self-improving loops require a signal that cannot be gamed. It has been argued that an LLM judge should yield to
a deterministic verification layer \cite{wahi2026llmasajudge}; test suites have been searched for as a denser
improvement signal \cite{cheng2026small}; a robot harness gates self-improvement on evidence
\cite{wang2026regenharness}; and poisoned benchmarks can contaminate self-modifying coding agents
\cite{roesner2026reflections}. Silent-failure detection by multi-layer quality gates has been studied with
fault injection in retrieval pipelines \cite{tanhaei2026silentfailure}, self-testing has been proposed as a
release gate for LLM applications \cite{maiorano2026automated}, and the non-determinism of flaky tests is a
long-standing obstacle to regression testing \cite{lam2020understanding}. We report that the deterministic gates
of an agent factory also failed, and by whom the failures were found.

\paragraph{Factories, automated science and oversight}
The term \emph{software factory} has recently been applied to agent swarms that coordinate through a shared
board such as GitHub, distilled from two industrial factories observed for 15 and 17 days, in which at least
5.6\,\% of issues marked as fixed were later found to have incorrect fixes \cite{jiang2026instant}. That work
designs and judges a factory by the changes it ships; this one measures how a factory's own rules and gates
change. Multi-agent frameworks cover the software life cycle end to end \cite{tawosi2025almas,ishibashi2024selforganized},
a tutorial surveys LLMs as collaborative development agents \cite{garg2026collaborative}, and agent-driven
audits of multi-agent specifications have been reported as a case study \cite{calboreanu2026iterative}. The AI
Scientist and its successor generate research papers end to end
\cite{lu2024the,yamada2025the}, PaperOrchestrator coordinates agents for scientific paper writing
\cite{yuan2026paperorchestrator}, and related agents extend the loop to science
\cite{lyu2026evoscientist,meng2026swarmrsi,wang2026rsiforge}. Workflows have been evolved with human
collaborators \cite{liu2026evolving}, a benchmark asks whether agents can build agents \cite{lu2026the}, and the
governance of persistent self-modifying agents has been analysed conceptually \cite{tallam2026layered}, as has
the feasibility of meaningful human oversight of increasingly autonomous systems \cite{holzinger2025is}. We found
no prior use of the term ``solution factory''; this is a statement about the phrase only. The novelty claimed is
the system: a factory whose own loop produced a non-software pipeline under human gating.

\section{The Factory and Its Improvement Loop}
\label{sec:factory}

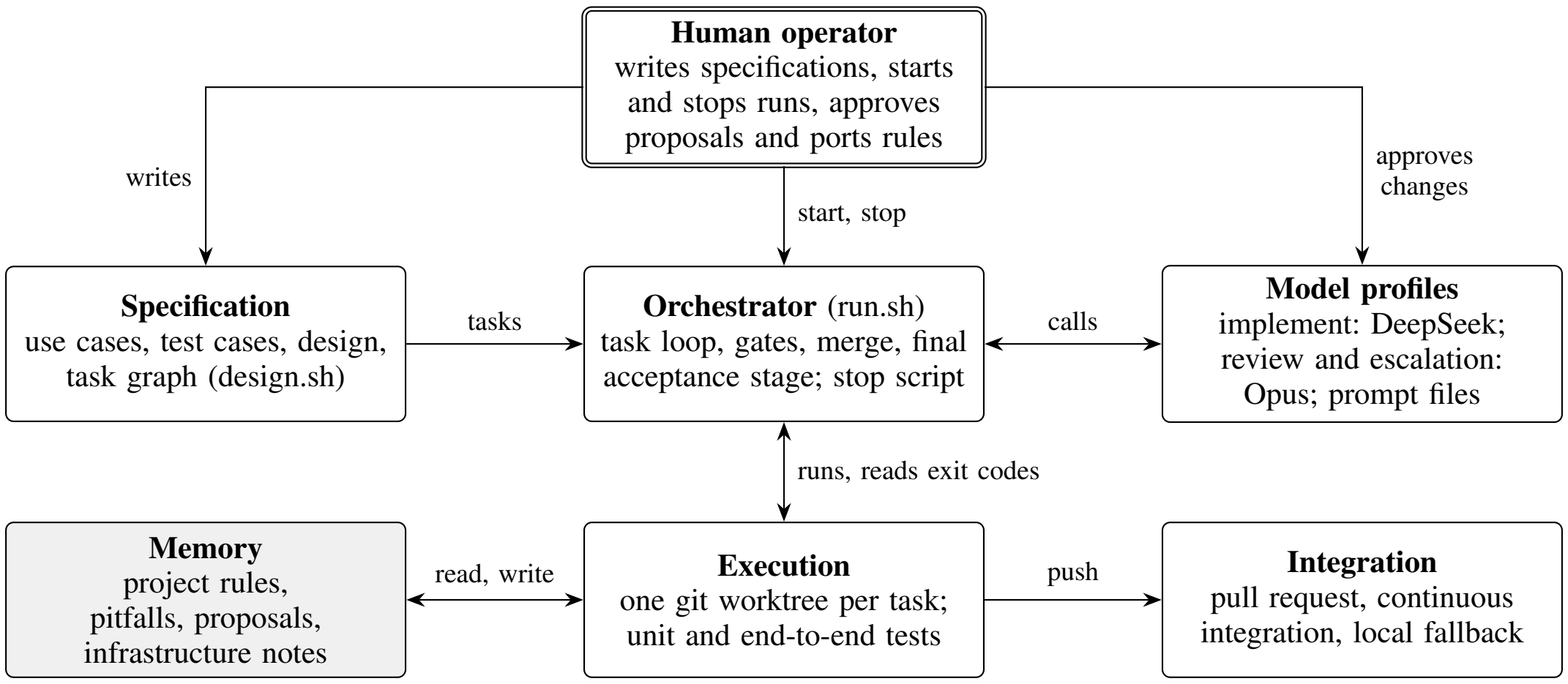
\begin{figure*}[t]
\centering
\begin{tikzpicture}[factory, compgrid]
\node[hum, comp] (human) {\textbf{Human operator}\\writes specifications, starts and stops runs,
  approves proposals and ports rules};
\node[box, comp, below=of human] (orch) {\textbf{Orchestrator} (run.sh)\\task loop, gates, merge, final
  acceptance stage; stop script};
\node[box, comp, left=of orch] (spec) {\textbf{Specification}\\use cases, test cases, design, task graph
  (design.sh)};
\node[box, comp, right=of orch] (models) {\textbf{Model profiles}\\implement: DeepSeek; review and
  escalation: Opus; prompt files};
\node[box, comp, below=of orch] (exec) {\textbf{Execution}\\one git worktree per task; unit and end-to-end
  tests};
\node[mem, comp, left=of exec] (memory) {\textbf{Memory}\\project rules, pitfalls, proposals,
  infrastructure notes};
\node[box, comp, right=of exec] (ci) {\textbf{Integration}\\pull request, continuous integration,
  local fallback};
\draw[->] (human) -- node[lbl, right] {start, stop} (orch);
\draw[->] (human.west) -| node[lbl, near end, left] {writes} (spec.north);
\draw[->] (human.east) -| node[lbl, near end, right] {approves\\changes} (models.north);
\draw[->] (spec) -- node[lbl, above] {tasks} (orch);
\draw[<->] (orch) -- node[lbl, above] {calls} (models);
\draw[<->] (orch) -- node[lbl, right] {runs, reads exit codes} (exec);
\draw[<->] (memory) -- node[lbl, above] {read, write} (exec);
\draw[->] (exec) -- node[lbl, above] {push} (ci);
\end{tikzpicture}
\caption{Components of the factory. Double borders mark the human role, the shaded box the memory files.
Arrows show calls and data flow.}
\label{fig:components}
\end{figure*}

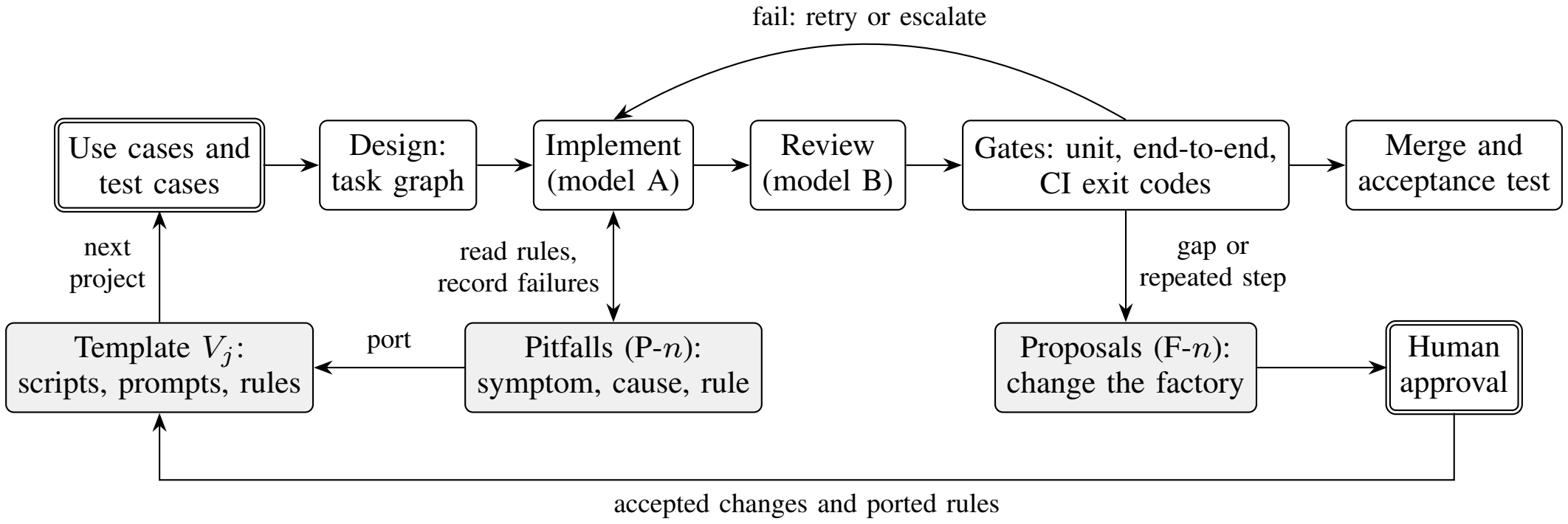
\begin{figure*}[t]
\centering
\begin{tikzpicture}[factory]
\node[hum] (spec) {Use cases and\\test cases};
\node[box, right=of spec] (design) {Design:\\task graph};
\node[box, right=of design] (impl) {Implement\\(model A)};
\node[box, right=of impl] (rev) {Review\\(model B)};
\node[box, right=of rev] (gates) {Gates: unit, end-to-end,\\CI exit codes};
\node[box, right=of gates] (merge) {Merge and\\acceptance test};
\draw[->] (spec) -- (design);
\draw[->] (design) -- (impl);
\draw[->] (impl) -- (rev);
\draw[->] (rev) -- (gates);
\draw[->] (gates) -- (merge);
\draw[->] (gates.north) to[bend right] node[lbl, above] {fail: retry or escalate} (impl.north);
\node[mem, below=of spec] (tmpl) {Template $V_j$:\\scripts, prompts, rules};
\node[mem, below=of impl] (pit) {Pitfalls (P-$n$):\\symptom, cause, rule};
\node[mem, below=of gates] (prop) {Proposals (F-$n$):\\change the factory};
\node[hum, below=of merge] (human) {Human\\approval};
\draw[<->] (impl) -- node[lbl, left] {read rules,\\record failures} (pit);
\draw[->] (gates) -- node[lbl, right] {gap or\\repeated step} (prop);
\draw[->] (prop) -- (human);
\draw[->] (pit) -- node[lbl, above] {port} (tmpl);
\coordinate[below=\tkdrop of human.south] (hdown);
\draw[->] (human.south) -- (hdown) -| node[lbl, near start, below] {accepted changes and ported rules}
  (tmpl.south);
\draw[->] (tmpl) -- node[lbl, left] {next\\project} (spec);
\end{tikzpicture}
\caption{Workflow of the factory and its improvement loop. Double borders mark human steps, shaded boxes the
memory files. Inside a project the agents read and write pitfalls and proposals; only the human changes the
template, from which the next project is created.}
\label{fig:arch}
\end{figure*}

The software factory is a set of shell scripts, prompt files and memory files that turns a written
specification into merged code. Fig.~\ref{fig:components} shows its components and Fig.~\ref{fig:arch} the
workflow through them. A human writes use cases and test cases; a design script splits them into tasks with
dependencies; an orchestrator script then runs each task in its own git worktree. One model implements the
task, a second model reviews it, and the task is merged only when the unit tests, the end-to-end tests and the
continuous-integration run exit with code zero. A final stage runs user-acceptance tests on the merged result.
An agent's report that a task is complete never counts as success; only exit codes do.

Three memory files sit beside the code. \emph{Pitfalls} are numbered entries (P-$n$), each with the
symptom, the root cause and a rule. Every agent must read them before it starts work. \emph{Factory
improvements} are numbered proposals (F-$n$) to change the factory itself, which a human approves or
rejects. \emph{Infrastructure notes} record problems with the network, limits or tools. Agents running
inside the loop may write pitfalls and proposals, but they may not edit the orchestrator or the project
rules; only the human applies an approved proposal.

The improvement operator works across projects. A failure in one project becomes a pitfall there. When the
human judges a lesson general, it is ported into the shared template, from which later projects are
created, and from there into other running projects. Fig.~\ref{alg:loop} states the loop. We index
template versions by project: $V_j$ is the template after project $p_j$ (in practice the template is updated
several times during a project; the algorithm merges these into one step). $R(p_j)$ is the set of rules in
force in project $p_j$, and $R_j$ is the rule component of the template $V_j$ (its pitfalls file), so a
project created from $V_j$ starts with $R_j$. The map $\mathrm{port}$ is a filter,
$\mathrm{port}(R(p_j)) \subseteq R(p_j)$, keeping the rules the human judges general. Definition~\ref{def:bounded} below gives the
template update exactly.

\begin{figure}[t]
\hrule\smallskip
{\small\noindent
\textbf{Input:} template $V_0$, rule set $R_0$, projects $p_1, p_2, \dots$ in time order.\\
\textbf{for} $j = 1, 2, \dots$ \textbf{do}\\
\hspace*{1em}create $p_j$ from template $V_{j-1}$ with rules $R(p_j) \leftarrow R_{j-1}$\\
\hspace*{1em}\textbf{for} each task in $p_j$ \textbf{do}\\
\hspace*{2em}implement; review must approve; merge only if unit,\\
\hspace*{2em}end-to-end and CI runs all exit with code $0$\\
\hspace*{2em}\textbf{on} a failure: write pitfall $r$ (symptom, cause, rule)\\
\hspace*{3em}$R(p_j) \leftarrow R(p_j) \cup \{r\}$\\
\hspace*{2em}\textbf{on} a repeated manual step or a gate gap: add proposal $f$ to $F_j$\\
\hspace*{1em}human reviews proposals $F_j$; accepts $A_j \subseteq F_j$\\
\hspace*{1em}$R_j \leftarrow R_{j-1} \cup \mathrm{port}(R(p_j))$\\
\hspace*{1em}$S_j \leftarrow S_{j-1} \oplus A_j$;\quad $V_j \leftarrow (S_j, R_j)$
\par}\smallskip\hrule
\caption{The bounded improvement loop.}
\label{alg:loop}
\end{figure}

\section{An Illustrative Formal Model}
\label{sec:model}

The model below is illustrative. It states, under explicit assumptions, what a rule can and cannot do to the
recurrence of a failure. It is not evidence for any claim: the claims are tested only on data in
Section~\ref{sec:results}. Table~\ref{tab:notation} collects the notation.

\begin{table}[t]
\centering
\caption{Notation.}
\label{tab:notation}
\resizebox{\columnwidth}{!}{%
\begin{tabular}{ll}
\toprule
Symbol & Meaning \\
\midrule
$p_j$ & the $j$-th project, in order of creation \\
$V_j = (S_j, R_j)$ & template after $p_j$: scripts and prompts, rules \\
$R(p_j)$ & rules in force in project $p_j$ \\
$F_j$, $A_j$ & proposals written in $p_j$; those a human accepted \\
$\mathrm{port}(\cdot)$ & filter selecting rules judged general \\
$K$, $k$ & set of failure classes; one class \\
$\lambda_k$ & rate of class-$k$ failures before a rule (per project-week) \\
$d_k$ & detection probability of the rule for class $k$ \\
$T$ & length of an observation window (project-weeks) \\
\bottomrule
\end{tabular}}
\end{table}

\begin{definition}[Failure classes and rules]
\label{def:classes}
Let $K$ be a finite set of failure classes. Before a rule for class $k \in K$ exists, failures of class $k$
occur in a project as a Poisson process with rate $\lambda_k > 0$ per project-week. A rule $r_k$ for class
$k$ is either \emph{check-backed} (it names a mechanical check such as a command, a test-harness case or a
gate) or \emph{prose-only}. Once $r_k$ is in force, each failure of class $k$ that would have occurred is
prevented with probability $d_k \in [0,1]$: the prevention marks are i.i.d.\ Bernoulli($d_k$) and
independent of the Poisson process of would-be failures. We call $d_k$ the detection
probability of the rule.
\end{definition}

\begin{definition}[Bounded self-improvement]
\label{def:bounded}
Write a template as a pair $V = (S, R)$ of its scripts and prompts $S$ and its rules $R$. A factory version
$V_j = (S_j, R_j)$ is \emph{bounded} with respect to $V_{j-1}$ if $S_j = S_{j-1} \oplus A_j$, where $A_j$ is a
finite set of written proposals that a human accepted and $\oplus$ applies them to $S$, and
$R_j = R_{j-1} \cup \mathrm{port}(R(p_j))$. These two equations are the only changes; no agent changes $S$
outside $A_j$. We further assume that the rule set grows
monotonically, $R_{j-1} \subseteq R_j$: this is an assumption about how the factory is run (agents may add
rules but not remove them), not a consequence of the definition.
\end{definition}

\begin{proposition}[Rule thinning]
\label{prop:thinning}
Assume the definitions above, with $\lambda_k$ and $d_k$ constant over the observation window. After the rule
$r_k$ comes into force, the failures of class $k$ that still occur form a Poisson process with rate
$(1-d_k)\lambda_k$. In particular, over a window of $T$ project-weeks the expected number of recurrences is
$(1-d_k)\lambda_k T$, and the probability of observing none is $e^{-(1-d_k)\lambda_k T}$.
\end{proposition}

\begin{proof}
Let $N(t)$ be the Poisson process of would-be failures with rate $\lambda_k$. Mark each event independently
as \emph{prevented} with probability $d_k$ and \emph{occurring} with probability $1-d_k$. By the thinning
property of the Poisson process, the occurring events form a Poisson process $N_o(t)$ with rate
$(1-d_k)\lambda_k$. To see this, fix an interval of length $T$ and condition on $N(T)=n$. The number of
occurring events is then binomial with parameters $n$ and $1-d_k$. Summing over $n$, using
$\binom{n}{m}/n! = 1/(m!\,(n-m)!)$ and substituting $j = n-m$,
\begin{align*}
\Pr[N_o(T)=m] &= \sum_{n \ge m} e^{-\lambda_k T}\frac{(\lambda_k T)^n}{n!}
                 \binom{n}{m}(1-d_k)^m d_k^{\,n-m} \\
&= e^{-\lambda_k T}\frac{((1-d_k)\lambda_k T)^m}{m!}\sum_{j \ge 0}\frac{(d_k\lambda_k T)^j}{j!} \\
&= e^{-(1-d_k)\lambda_k T}\frac{((1-d_k)\lambda_k T)^m}{m!},
\end{align*}
since $\sum_{j \ge 0}(d_k\lambda_k T)^j/j! = e^{d_k\lambda_k T}$; this is the Poisson law with mean
$(1-d_k)\lambda_k T$. For disjoint intervals, the counts of would-be failures are independent (Poisson
increments), and the marks are independent of these counts and of each other, so the counts of occurring
failures on disjoint intervals are independent as well. Hence $N_o$ is a Poisson process. The expectation and the
probability of $m=0$ follow from the Poisson law.
\end{proof}

The proposition says that a rule with $d_k<1$ lowers, but does not remove, the chance of a recurrence. It also
says why a raw comparison of rule types can mislead, which the next statement makes exact.

\begin{proposition}[Confounding by failure frequency]
\label{prop:confound}
Let $c$ be a check-backed rule and $p$ a prose-only rule with $0 \le d_p < d_c < 1$ (so both rules are
imperfect). Under
Proposition~\ref{prop:thinning}, with $\lambda_c, \lambda_p > 0$ as in Definition~\ref{def:classes}, the check-backed rule has
the higher expected recurrence rate (expected recurrences per project-week) if and only if
\[
\frac{\lambda_c}{\lambda_p} > \frac{1-d_p}{1-d_c}.
\]
\end{proposition}

\begin{proof}
By Proposition~\ref{prop:thinning}, the expected recurrence rates are $(1-d_c)\lambda_c$ and
$(1-d_p)\lambda_p$. Both $1-d_c$ and $\lambda_p$ are positive, so
$(1-d_c)\lambda_c > (1-d_p)\lambda_p$ holds if and only if
$\lambda_c/\lambda_p > (1-d_p)/(1-d_c)$.
\end{proof}

Proposition~\ref{prop:confound} is the reason the comparison of check-backed and prose-only rules in
Section~\ref{sec:results} cannot be read as a test of $d_c > d_p$ alone. If authors attach checks to the
failure classes that occur most often, a better rule can still show more recurrences. Separating the two
needs class-matched rates, which our data do not provide.

\section{Study Design}
\label{sec:design}

\paragraph{Data}
We read \nRepos{} git repositories of the factory lineage: the template, and the projects created from it or
ported into it (Table~\ref{tab:projects}). Each was read at its last commit on the main branch dated no
later than \cutoffDate{}; Table~\ref{tab:snapshot} lists those commits, so anyone can check out the same data and rerun
the analysis. Three sources per repository: the pitfalls file, the factory-improvements file and the commit
log. Scripts parse them, and every count in the tables and the text is theirs.

\begin{table}[t]
\centering
\caption{The repositories. Days: from the first to the last commit read, both included. Tasks: distinct task
identifiers merged on the main branch. Native P: pitfall rules written in that repository; Ported P: copies
received from another repository; F: improvement proposals recorded in that repository's own file.}
\label{tab:projects}
\resizebox{\columnwidth}{!}{\begin{tabular}{lrrrrrrr}
\toprule
Repository & Start & Days & Commits & Tasks & Native P & Ported P & F \\
\midrule
template & 2026-11-08 & 18 & 44 & 0 & 5 & 41 & 8 \\
stock\_game & 2026-11-08 & 20 & 229 & 60 & 33 & 0 & 24 \\
new\_game & 2026-11-08 & 15 & 142 & 21 & 12 & 8 & 20 \\
quant\_home & 2026-11-12 & 16 & 212 & 25 & 36 & 4 & 20 \\
paper\_agent & 2026-11-22 & 5 & 58 & 15 & 17 & 25 & 15 \\
learning\_agent & 2026-11-25 & 3 & 24 & 6 & 1 & 45 & 9 \\
\midrule
Total &  &  & 709 & 127 & 104 & 123 & 96 \\
\bottomrule
\end{tabular}
}
\end{table}

\begin{table}[t]
\centering
\caption{Commit of each repository used for every number in the paper.}
\label{tab:snapshot}
\begin{tabular}{ll}
\toprule
Repository & Commit at cutoff \\
\midrule
template & \texttt{cf7b22e12b77} \\
stock\_game & \texttt{90a69634207b} \\
new\_game & \texttt{f994bde81de3} \\
quant\_home & \texttt{0a7a3f5ff3c9} \\
paper\_agent & \texttt{6d53d2d22511} \\
learning\_agent & \texttt{581023abb6af} \\
\bottomrule
\end{tabular}

\end{table}

\paragraph{Coding the rules}
An entry is \emph{ported} if its title says it was ported from another repository or its ``Seen in'' line
starts with another repository's name; otherwise it is \emph{native}. Native entries are the unique rules; a
ported entry is a copy. A rule is \emph{check-backed} if its rule text names a mechanical check (a command, a
test-harness case or a gate) and \emph{prose-only} otherwise. We code this on the rule text in the commit that
first added it, so that a check added later, perhaps because the failure came back, cannot bias the split.

\paragraph{Failure classes}
Each unique rule is assigned to a failure class by keyword patterns on its title, applied in a fixed order
(platform and shell; third-party services; user interface; tests and gates; specification and design;
orchestration and process; runtime and environment; data and content); a title that matches none is classed as
``Other''. The coding is deterministic, and the full rule-to-class list is published with the scripts so that
it can be audited or recoded.

\paragraph{Recurrence}
A rule \emph{recurred} if its entry carries a dated note of a new occurrence (``Seen again'', ``Repeated'',
``Broken again'') or its title says the failure happened ``again''. A note in a ported copy counts for the
receiving repository only if the same note is not on the original rule. This measure relies on what the
agents and the operator wrote down. A recurrence nobody recorded is invisible, so the rates are lower bounds.

\paragraph{Tests}
Each claim has a test fixed before the analysis. For \cA{}, the test is whether most recurrences are of rules
already ported into the repository where they recur; if so, the claim fails. A second test, before-versus-after
porting rates per failure class, was planned but needs every failure matched to a class across repositories by
hand, and we did not do it, so \cA{} is only partly tested. \cB{} is refuted by no difference, or by check-backed
rules recurring as often as prose-only rules or more. For \cC{}, the harness must not have found the gate
defects itself, nor have been able to: we check that it passed on the commit where each defect was then found.
\cD{} holds only if proposals per delivered task fall strictly from one project to the next, in the order the
projects started. Improvement proposals stand in for human decisions; gate approvals and
escalations are not counted. With no ungated factory to compare against, \cD{} is descriptive, not causal.

\section{Results}
\label{sec:results}

\subsection{Descriptive overview}
\begin{figure*}[t]
\centering
\includegraphics[width=\textwidth]{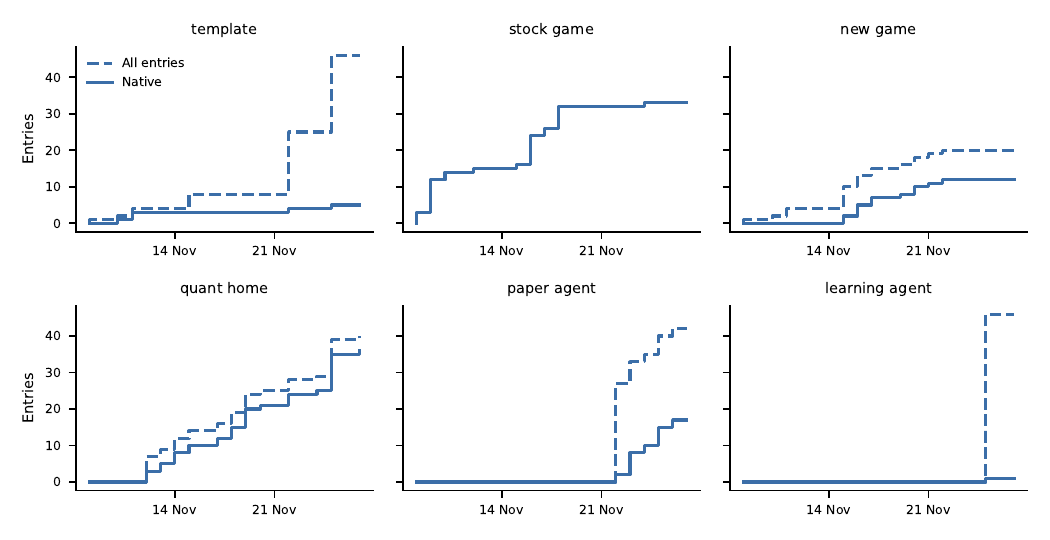}
\caption{Cumulative pitfall entries per repository by the date each entry was first committed. Solid: native
rules; dashed: all entries including ported copies.}
\label{fig:timeline}
\end{figure*}

\begin{figure}[t]
\centering
\includegraphics[width=0.9\columnwidth]{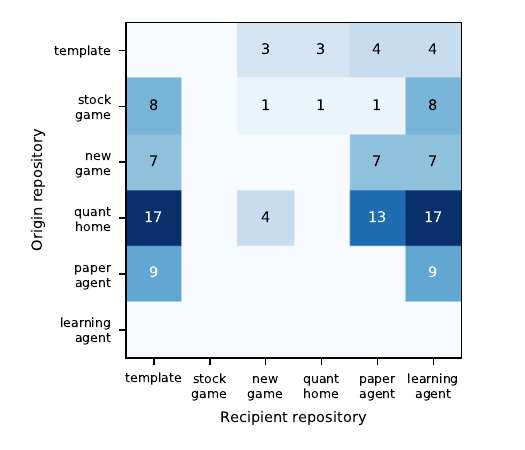}
\caption{Propagation of rules between repositories: number of entries in each recipient (column) ported from
each origin (row).}
\label{fig:flow}
\end{figure}

Fig.~\ref{fig:timeline} shows how the rule base of each repository grew between \firstRuleDate{} and
\lastRuleDate{}. Native rules accumulate gradually during active development, whereas ported entries arrive in
steps, each step corresponding to a porting session in which the operator copied a batch of lessons. The youngest
repository consists almost entirely of ported entries, and in the second youngest they outnumber the native
ones. Fig.~\ref{fig:flow} shows the direction of
propagation: \nFlows{} origin--recipient pairs carry at least one rule, and the largest flows, of
\maxFlow{} entries each, run from \maxFlowPairs{}. The template therefore acts as a hub: it collects
lessons from production projects and redistributes them to new ones, and it is itself the origin of
\nOriginTemplate{} ported entries.

Table~\ref{tab:taxonomy} classifies the unique rules. \topClass{} is the largest class, with \nTopClass{} of
\nUniqueRules{} rules, which reflects how much of the operator's effort went into the machinery that decides
whether work is accepted. Recorded recurrences fall into \nRecurClasses{} classes. Rules that could not be
assigned to a class (\nOtherClass{}) are mostly domain-specific defects of individual products.

\begin{table}[t]
\centering
\caption{Failure classes of the unique rules, with the share that named a check when first written and the
number of recorded recurrences.}
\label{tab:taxonomy}
\begin{tabular}{lrrr}
\toprule
Failure class & Rules & Check-backed (\%) & Recurred \\
\midrule
Platform and shell & 11 & 64 & 2 \\
Third-party services & 7 & 57 & 0 \\
User interface & 4 & 50 & 0 \\
Tests and gates & 38 & 63 & 2 \\
Specification and design & 10 & 40 & 0 \\
Orchestration and process & 6 & 50 & 0 \\
Runtime and environment & 8 & 50 & 1 \\
Data and content & 8 & 38 & 0 \\
Other & 12 & 33 & 0 \\
\midrule
Total & 104 & 53 & 5 \\
\bottomrule
\end{tabular}

\end{table}

\subsection{\rqA{}: transfer across projects (\cA{})}
\begin{figure}[t]
\centering
\includegraphics[width=\columnwidth]{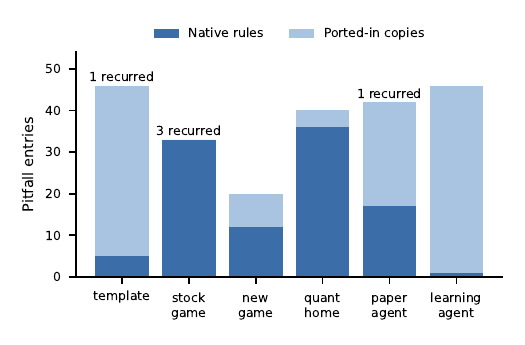}
\caption{Pitfall entries per repository, native and ported in, with recurrences marked.}
\label{fig:rules}
\end{figure}

Of \nUniqueRules{} unique rules, \nRecur{} recorded a recurrence (\pctRecur{}\,\%), and in every case the
recurrence occurred in the repository in which the rule had been written (Table~\ref{tab:transfer}). None of the
\nPortedCopies{} ported copies recorded a recurrence in its receiving repository; \nCopiedNotes{} copies carried
a recurrence note that had been copied together with the rule, and these were not counted.
Fig.~\ref{fig:rules} shows the distribution of rules across repositories. The disproval test that could be run
is passed: recurrences are not predominantly of ported rules. The planned before-and-after comparison was not
performed, so \cA{} has at most weak support. Two selection effects also favour the result: rules are usually
ported after they have proved useful, and receiving repositories are younger, leaving less time for a failure
to return.

\begin{table}[t]
\centering
\caption{Where rules recurred (\cA{}).}
\label{tab:transfer}
\begin{tabular}{lrr}
\toprule
Where the rule recurred & Rule copies & Recurrences \\
\midrule
Origin repository (rule written there) & 104 & 5 \\
Recipient repository (rule ported in) & 123 & 0 \\
\bottomrule
\end{tabular}

\end{table}

\subsection{\rqB{}: check-backed and prose-only rules (\cB{})}
All \nRecur{} recorded recurrences concern rules that named a check when first written: \nRecurCheck{} of
\nCheckRules{} check-backed rules (\pctRecurCheck{}\,\%) against \nRecurProse{} of \nProseRules{} prose-only
rules (Table~\ref{tab:ruletype}). \cB{} therefore fails its test. Table~\ref{tab:recurrences} lists each
recurrence so that the coding can be verified. The recurring failures are of one kind: a stalled end-to-end
test, an acceptance-report format, a socket that did not reconnect, and two cases of editing files through the
Windows shell. These are frequent, platform-level failures, and the two classes that hold most of them, platform
and shell and tests and gates, also have the highest check-backed shares in Table~\ref{tab:taxonomy}. This is the pattern that Proposition~\ref{prop:confound}
describes: when checks are attached preferentially to frequent failure classes, a more effective rule type can
show more recurrences. With at most a handful of recurrences per class, a class-matched comparison has no
statistical power, and the data cannot separate the two effects.

\begin{table}[t]
\centering
\caption{Recurrence by rule type, coded on the rule as first written (\cB{}).}
\label{tab:ruletype}
\begin{tabular}{lrrr}
\toprule
Rule type (as first written) & Rules & Recurred & Share (\%) \\
\midrule
Check-backed & 55 & 5 & 9.1 \\
Prose only & 49 & 0 & 0.0 \\
\midrule
All unique rules & 104 & 5 & 4.8 \\
\bottomrule
\end{tabular}

\end{table}

\begin{table}[t]
\centering
\caption{Every recorded recurrence, with the note that records it.}
\label{tab:recurrences}
\resizebox{\columnwidth}{!}{\begin{tabular}{llp{0.62\columnwidth}}
\toprule
Repo & Rule & Recurrence note (truncated) \\
\midrule
template & P-002 & - Seen again 2026-11-25 (porting F-041..F-045): a Python heredoc... \\
stock\_game & P-003 & - Seen again in T28, after the socket\_timeout=30 fix: attempts 1–2... \\
stock\_game & P-008 & - Broken again 2026-11-11: the human UAT section added DEF-1 under a... \\
stock\_game & P-025 & - Fixed 2026-11-23 (seen again on the first Lighthouse staging... \\
paper\_agent & P-039 & title: Windows shell quoting broke commits and one-liners again \\
\bottomrule
\end{tabular}
}
\end{table}

\subsection{\rqC{}: verification of the gates (\cC{})}
Table~\ref{tab:gates} lists \nGateDefects{} defects in the factory's own gates, that is, in the scripts, and the
prompts that drive them, that decide whether a task passed, failed or should pause. By the harness we mean the factory's own offline unit and end-to-end suites that exercise these scripts; acceptance
testing of the factory is a separate activity and is listed by who performed it. The harness detected \nGateByHarness{} of the defects. An
independent second agent, commissioned to perform acceptance testing of the factory, detected
\nGateBySecondAgent{}; a production run exposed \nGateByProduction{}; and an earlier acceptance test of the template, run outside the harness, found
\nGateByUAT{}. On the commits at which the second agent found its defects, the unit and end-to-end suites of the harness had
passed. One
defect was introduced by the fix for another and was again found by the second agent. The defects follow a
recurring pattern: a gate tested only on single-item lists, a gate whose input normalisation differed from that
of its test, and a gate that accepted a test identifier occurring in a comment. \cC{} passes its test in this
factory. Deterministic gates have been argued to be necessary \cite{wahi2026llmasajudge}; here they were not
sufficient, and an independent tester found what their own harness missed.

\begin{table}[t]
\centering
\caption{Defects in the factory's own gates and who found them (\cC{}).}
\label{tab:gates}
\resizebox{\columnwidth}{!}{\begin{tabular}{lp{0.48\columnwidth}l}
\toprule
ID & Defect & Found by \\
\midrule
DEF-1 & CI-fix loop fired on the task's own gate failures & second agent \\
DEF-2 & push/PR failure failed the task instead of pausing & second agent \\
DEF-3 & machine fault recorded as a flaky test & second agent \\
DEF-4 & final report hid lost auth or stuck CI & second agent \\
DEF-5 & wrong cross-references in two prompts & second agent \\
DEF-6 & UAT-ID gate counted an ID in a comment & second agent \\
DEF-7 & final e2e skip pattern too wide & second agent \\
DEF-8 & empty task branch paused forever (introduced by the DEF-2 fix) & second agent \\
P-001 & CRLF from jq: 15 of 16 UAT IDs reported missing & production run \\
P-004 & verdict whitespace strip let PASS hide open defects & UAT \\
\bottomrule
\end{tabular}
}
\end{table}

\subsection{\rqD{}: human effort per task (\cD{})}
\begin{figure}[t]
\centering
\includegraphics[width=\columnwidth]{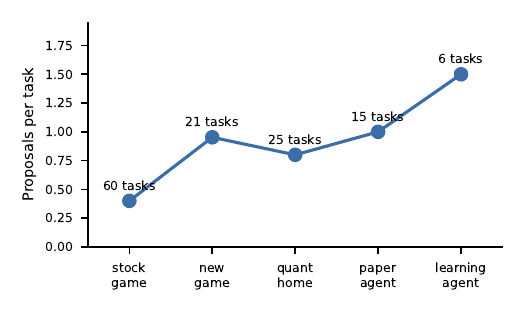}
\caption{Improvement proposals per delivered task, projects in order of creation (\cD{}).}
\label{fig:decisions}
\end{figure}

Fig.~\ref{fig:decisions} and Table~\ref{tab:decisions} report proposals per delivered task, with projects in
order of creation. From \ratioFirst{} (\nameFirst{}) the ratio rose to \ratioLast{} (\nameLast{}), increasing
in \nIncreases{} of \nSteps{} steps; \cD{} fails its test. Of the \nFInRatio{} proposals made in repositories that delivered tasks (the template, which delivers none,
is excluded), \nApplied{}
(\pctApplied{}\,\%) were applied and \nRejected{} were rejected; the remainder await a decision. Later projects
were smaller, and one of them, the paper pipeline of Section~\ref{sec:solution}, changed the factory as it was
built, so that each task touched the factory more often. No proposal was rejected in the observation window, so
these data show the human gate approving and deferring proposals, not filtering them. Even so, the expectation that the loop would require less of
its human operator as it learned is not supported. This result is descriptive: without an ungated factory for
comparison, it records what the loop asked of its operator, not why.

\begin{table}[t]
\centering
\caption{Improvement proposals and their status per project (\cD{}). Open: not yet decided.}
\label{tab:decisions}
\resizebox{\columnwidth}{!}{\begin{tabular}{lrrrrrr}
\toprule
Repository & Tasks & F & Applied & Open & Rejected & F/task \\
\midrule
stock\_game & 60 & 24 & 18 & 6 & 0 & 0.40 \\
new\_game & 21 & 20 & 2 & 18 & 0 & 0.95 \\
quant\_home & 25 & 20 & 2 & 18 & 0 & 0.80 \\
paper\_agent & 15 & 15 & 10 & 5 & 0 & 1.00 \\
learning\_agent & 6 & 9 & 4 & 5 & 0 & 1.50 \\
\bottomrule
\end{tabular}
}
\end{table}

\section{From Software Factory to Solution Factory}
\label{sec:solution}

Not every repository in Table~\ref{tab:projects} is an ordinary software product. One, paper\_agent, holds a
pipeline whose purpose is to write research papers, with human approval gates between its steps. It was built
the same way as a game and a trading tool: from use cases, test cases and
tasks, through the same gates. Building it did change the factory, but only through the usual path: the project
wrote its own improvement proposals, and the human ported the accepted ones into the template
(Table~\ref{tab:projects} lists its proposals). The rules from earlier projects came along.

This paper is the pipeline's output, run on the factory template itself. It is one instance and we claim no
generality from it. It does show a direction. A factory that can build the pipeline for a new kind of work,
and that carries its lessons into that pipeline, is no longer only a software factory. We call it a
\emph{solution factory}: the product is a working, gated process for a class of problems, not just code. For
our one factory, the results above suggest what this costs: recorded recurrences stayed where rules were
written (\cA{}, weak support), its own gates needed an independent tester (\cC{}), and proposals per task did
not fall (\cD{}, descriptive).

\section{Discussion}
\label{sec:discussion}

\subsection{Implications}
\paragraph{For RSI research}
Benchmark-scored self-rewriting optimises a scalar. The loop studied here optimises a written rule set that can
be read, audited and rejected, and this property is what made the negative results observable: a rule that
failed to prevent its failure records the fact in its own entry, and a gate that failed silently is listed with
the agent that detected it. For the safety of self-improving systems, auditability of this kind may matter as
much as the rate of improvement, since every change to the factory carries an author, a rationale and an
approver.

\paragraph{For practitioners}
Three practices follow from the results. First, a gate that decides acceptance should be tested on inputs of
more than one item and with surrounding whitespace and platform-specific line endings, since these were the
recurring causes of silent gate failure. Second, the verifiers of an agent pipeline benefit from an
independent tester that did not write them. Third, the human cost of a gated loop should be monitored
explicitly; in this factory it did not decline by itself.

\subsection{Threats to validity}
\paragraph{Construct validity}
Because recurrence is self-reported, the true rate may exceed the measured one by an unknown amount.
Check-backed is a property of a rule's text; the model of Section~\ref{sec:model} would instead require a
measured detection probability. Keyword-based assignment of failure classes may misclassify individual rules. Proposals stand in for human decisions and omit gate approvals and escalations, and delivered
tasks are counted from commit subjects rather than merged pull requests.

\paragraph{Internal validity}
No ungated factory is available for comparison, so \cD{} is descriptive. The \cB{} comparison may be confounded
by failure frequency (Proposition~\ref{prop:confound}). The \cA{} test passed, but the planned before-and-after
comparison per failure class was not performed. Who found each gate defect was coded by hand, with each code
tied to a line in the repository; no second rater checked the coding.

\paragraph{External validity}
One operator ran all \nRepos{} repositories over \nDaysObserved{} days, on one operating system, with the same two
model families. The repositories share a lineage and are not independent samples. A conflict of interest must
also be disclosed: a pipeline built by the factory wrote this paper, and the paper's author designed the
factory. As a safeguard, every number is produced by a script that reads the repositories at fixed commits.

\section{Conclusion}
\label{sec:conclusion}

We measured a bounded, human-gated self-improvement loop in an LLM software factory across \nRepos{}
repositories. None of \nPortedCopies{} ported rules recorded a recurrence in its new repository, while
\nRecur{} of \nUniqueRules{} rules recorded one where they were written; with self-reported recurrence and no
before-and-after comparison, this is weak support for transfer (\cA{}). Check-backed rules recurred more often
than prose rules, \pctRecurCheck{}\,\% against \pctRecurProse{}\,\% (\cB{}, not supported), a comparison that a
simple model shows can be confounded by how often the guarded failure occurs. The deterministic gates that guard
the loop had defects their own harness missed: of \nGateDefects{} gate defects, the harness found
\nGateByHarness{} and an independent second agent found \nGateBySecondAgent{} (\cC{}, supported in this one
factory). Proposals per delivered task did not fall; they went from \ratioFirst{} to \ratioLast{} (\cD{}, not
supported, descriptive). The same loop produced a non-software pipeline, the one that wrote this paper (\cE{},
a position).

In this one factory, bounded RSI recorded no recurrence of the lessons it ported, and it did
not run itself: its verifiers needed verifying, and it kept asking for its human. Future work should compare
gated and ungated factories, add operators, and replace self-reported recurrence with failure classes matched
across repositories.

\section*{Acknowledgments}
AI-use disclosure. This paper was produced with the paper\_agent pipeline described in
Section~\ref{sec:solution}. Claude Code (Anthropic, model Claude Opus) proposed the topic, ran the literature
searches, drafted the claims and the outline, wrote the analysis scripts, the formal model with its proofs, and
the text of every section. DeepSeek reviewed the outline, each script, the proofs and each section. The author
chose the topic, approved the literature, the claims and the target venue at human gates, decided how to report
the negative results, and is responsible for the final text. All numbers come from scripts run on the
repositories at the commits listed in Table~\ref{tab:snapshot}.
\bibliographystyle{IEEEtran}
\bibliography{references_ieee}
\end{document}